\documentclass[preprintnumbers,nofootinbib,a4paper,superscriptaddress]{revtex4-1}

\usepackage{amsmath,latexsym,amssymb,amsfonts}
\usepackage{float} 
\usepackage{graphicx}
\usepackage{epstopdf}
\usepackage{graphics}
\usepackage{caption}
\usepackage{subfigure}
\usepackage{bm}
\usepackage{color}
\usepackage{hyperref}
\usepackage[dvips]{psfrag}

\usepackage{soul}  
\usepackage{xcolor}
\setstcolor{red}

\begin{document}

\title{Constraints on dark energy models from the Horndeski theory}

\author{Bilguun Bayarsaikhan}
\affiliation{Institute of Physics \& Technology,
Mongolian Academy of Science, Ulaanbaatar 13330, Mongolia}
\author{ Seoktae Koh}
\affiliation{Department of Science Education, Jeju National University, Jeju, 63243, Korea}
\author{Enkhbat Tsedenbaljir}
\affiliation{Institute of Physics \& Technology,
Mongolian Academy of Science, Ulaanbaatar 13330, Mongolia}
\author{Gansukh Tumurtushaa}
\email[E-mail: ]{gansuh.mgl@gmail.com}
\affiliation{Department of Science Education, Jeju National University, Jeju, 63243, Korea}
\affiliation{Center for Theoretical Physics of the Universe, Institute for Basic Science (IBS), Daejeon 34051, Korea}
\preprint{CTPU-PTC-20-12} 
\begin{abstract}

In light of the cosmological observations, we investigate dark energy models from the Horndeski theory of gravity. In particular, we consider cosmological models with  the derivative self-interaction of the scalar field and the derivative coupling between the scalar field and gravity. We choose the self-interaction term to have an exponential function of the scalar field with both positive and negative exponents. For the function that has a positive exponent, our result shows that the derivative self-interaction term plays an important role in the late-time universe. On the other hand, to reproduce the right cosmic history, the derivative coupling between the scalar field and gravity must dominate during the radiation-dominated phase. However, the importance of such a coupling in the present universe found to be negligible due to its drastic decrease over time.  Moreover, the propagation speed of gravitational waves estimated for our model is within the observational bounds, and our model satisfies the observational constraints on the dark energy equation of state.

\end{abstract}

\maketitle

\section{Introduction}

The observations of Supernovae type Ia (Sn Ia)~\cite{Riess:1998cb, Perlmutter:1998np}  in the late 1990s have led to the discovery that our universe is expanding at an accelerated rate. An explanation for this accelerated expansion invokes the existence of a mysterious form of energy, \emph{i.e.,} dark energy~\cite{Perlmutter:1998np}. The independent measurements of Cosmic Microwave Background (CMB) temperature anisotropy~\cite{Spergel:2003cb, Hinshaw:2012aka, Ade:2013zuv, Ade:2015xua, Aghanim:2018eyx} and the Baryon Acoustic Oscillation (BAO)~\cite{Eisenstein:2005su} all support its existence. The observations favor $\omega_{DE} \simeq -1$, where $\omega_{DE}$ is the equation of state of dark energy. Although the origin of dark energy is still unknown, there have been suggested many models explain the present accelerating universe~\cite{DeFelice:2010aj}.
 
The most popular dark energy model is the cosmological constant $\Lambda$, the positive vacuum energy density in the  $\Lambda$CDM model of the universe.
The $\Lambda$CDM model, based on Einstein's theory of general relativity (GR), gives the robust description of the universe within a few handful parameters and hence regarded as the standard model in cosmology. However, the cosmological constant suffers from the so-called fine-tuning and coincidence problems~\cite{Carroll:2000fy, Sahni:1999gb}. Alternatives to the cosmological constant often involve additional degrees of freedom such as scalar, vector, and tensor fields. In this work, we focus on the scalar-tensor theories of gravity, where only one additional scalar degree of freedom is included.

A slowly rolling scalar field has been a popular candidate for inflation and dark energy. The potential energy density of such a field is undiluted by the expansion of the universe; hence it can act as the effective cosmological constant driving phases of acceleration. Therefore, models of inflation and dark energy based on scalar fields depend on a specific form of the scalar-field potential $V(\phi)$. However, it is generally known that it is difficult to realize the potential from the fundamental theory to fulfill the slow-roll conditions in inflation or to resolve the coincidence problems in dark energy models~\cite{Carroll:2000fy, Sahni:1999gb}. One of the criticisms about potential driven inflation and dark energy models is given
by the swampland conjectures in string theory. The swampland conjecture in the cosmological background put the constraints on the field range of a scalar field, $|\Delta \phi|/M_{pl}\leq d$ where $M_{pl}=1/\sqrt{8\pi G}$ is the reduced Planck mass, and on the slope of the potential of such fields, $M_{pl}|V'|/V>c$~\cite{Ooguri:2006in, Obied:2018sgi, Agrawal:2018own, Ooguri:2018wrx}. Here, $d$ and $c$ are positive constants of  $\sim \mathcal{O}(1)$ over a certain range for the scalar field. The implications of these conjectures have been studied for cosmic inflation~\cite{inflation} and dark energy~\cite{darkenergy, Heisenberg:2018yae, Heisenberg:2019qxz, Brahma:2019kch}. \emph{For example}, the second constraint implies that the model parameter $\lambda\sim\mathcal{O}(1)$ in $V(\phi)\sim e^{-\lambda \phi/M_{pl}}$.

Over the past decade, as a consistent ghost-free effective field theory, Horndeski theory of gravity~\cite{Horndeski:1974wa} has been studied extensively for scalar-field dark energy. Horndeski's theory of gravity is regarded as the most general scalar-tensor theory of gravity whose dynamics governed by second-order equations of motion. 
The action in the Horndeski theory can be written as~\cite{Deffayet:2011gz, Kobayashi:2011nu, Gao:2011vs}
\begin{eqnarray}\label{eq:Horndeski}
S = \int d^{4}x \sqrt{-g}\left( L_2+L_3+L_4+L_5 \right)+S_{m,r}\,,
\end{eqnarray}
where $S_{m,r}$ is the action for ordinary matter and radiation, excluding the scalar field, and 
{\small \begin{eqnarray}
L_2 &=& K(\phi, X)\,, \quad L_3 = G_3(\phi, X) \square\phi\,, \quad L_4 = G_4(\phi, X) R + G_{4,X}\left[\left(\square \phi\right)^2 - \left(\nabla_\mu\phi\nabla_\nu\phi\right)\left(\nabla^\mu\phi\nabla^\nu\phi\right)\right]\,,\\
L_5 &=& G_5(\phi, X) G_{\mu\nu}\left(\nabla^\mu\nabla^\nu\phi\right)-\frac16 G_{5,X}\left[(\square\phi)^3 -3 \square\phi \left(\nabla_\mu\phi\nabla_\nu\phi\right)\left(\nabla^\mu\phi\nabla^\nu\phi\right)+2 \left(\nabla^\mu\phi\nabla_\alpha\phi\right)\left(\nabla^\alpha\phi\nabla_\beta\phi\right)\left(\nabla^\beta\phi\nabla_\mu\phi\right)\right]\,,\nonumber
\end{eqnarray}} 
with $X = -\nabla_\mu\phi\nabla^\mu\phi/2$, $\square\phi = \nabla_\mu\nabla^\mu\phi$, and $G_{i,X}(\phi,X) = \partial G_i(\phi,X)/\partial X$ ($i=4,5$). The functions $K(\phi, X)$ and $G_i(\phi, X)\,\,(i=3,4,5)$ are arbitrary functions of $\phi$ and $X$. For $K=G_3=G_5=0$ and $G_4=M_{pl}^2/2$, the action reduces to that of GR. By employing special combinations of the independent functions $K(\phi, X)$ and $G_i(\phi, X)$, one can construct a broad spectrum of cosmological models from Eq.~(\ref{eq:Horndeski}). 

The advent of Multi-messenger astronomy has brought a new era in cosmological studies. In particular, the direct detections of gravitational waves (GWs) from a neutron star merger GW170817~\cite{TheLIGOScientific:2017qsa} and its associated electromagnetic counterpart GRB170817A~\cite{Monitor:2017mdv}  
allow us to constrain the GW speed with a remarkable precision:
\begin{align}\label{eq:const}
-3\times 10^{-15} \leq \frac{c_T}{c}-1\leq 7\times 10^{-16}\,,
\end{align}
where $c_T$ the sound speed for tensor perturbation and $c$ the speed of light. This bound indicates that the difference in propagation speed between light and gravitational waves to be less than about one part in $10^{15}$. In the Horndeski theory, the $c_T$ is expressed as~\cite{Kobayashi:2011nu}
\begin{align}\label{eq:ct2}
c_T^2 = \frac{G_4 - X \left(G_{5, \phi} + G_{5, X}\ddot{\phi}\right)}{G_4-2 X G_{4, X} + X\left( G_{5, \phi} - G_{5, X} H \dot{\phi} \right) }\,.
\end{align}
An immediate consequence of Eq.~(\ref{eq:ct2}) is that the propagation speed $c_T^2$ of GWs is independent of $K(\phi, X)$ and $G_3(\phi, X)$. 
The requirement of luminal propagation of GWs within this framework implies the condition: $G_{4, X}=0$ and $G_{5}=\text{\emph{const.}}$~\cite{Ezquiaga:2017ekz}. Cosmological models, based on the Horndeski theory, are ruled out unless this condition is satisfied. As we will shortly see, in our model, $G_{5}(\phi)$ is a function of a scalar field $\phi$, which may violate the above condition. Therefore, it is imperative for us to examine the propagation speed of GWs. 

Our purpose in the present work is to study the late-time dynamics of the universe for a subclass of the Horndeski theory in light of observational constraints, including bounds on the propagation speed of GWs. In particular, we consider cosmological models with the derivative self-interaction of the scalar field and its derivative coupling to gravity. Thus, the setup for models we investigate in this work is the following:
\begin{eqnarray}\label{eq:setup}
K(\phi, X) = X- V(\phi)\,, \qquad G_{3}(\phi,X) = \frac{\alpha}{M^3}\xi(\phi) X \,, \qquad G_4 = \frac{M_{pl}^2}{2}\,, \qquad G_5(\phi)=\frac{\beta}{2M^2}\phi\,,
\end{eqnarray}  
where $V(\phi)$ is the scalar-field potential, $\alpha$ and $\beta$ are dimensionless constants, and $M$ is a mass scale. This model has been employed previously for studying cosmic inflation in Ref.~\cite{Tumurtushaa:2019bmc} and yield results consistent with the CMB observations. For the setup, we investigate the dynamical evolution of both derivative self-interaction of the scalar field $G_{3}(\phi, X)$ and derivative coupling between the scalar field and gravity $G_5(\phi)$ and discuss their cosmological implications in the present universe in view of the observational constraints. To better understand their late-time behavior, we perform the so-called dynamical system analysis of the fixed points. 

The remainder of the paper is organized as follows. For the setup in Eq.~(\ref{eq:setup}), we derive the equations of motion in Sec.~\ref{sec:setup}. The dynamical system analyses of the fixed points is discussed in Sec.~\ref{sec:analysis} wherein we rewrite the background equations of motion in terms of dimensionless variables. In Sec.~\ref{sec:model}, we present our numerical results for an explicit model. We examine the propagation speed of GWs  for our model by using the observation bounds in Sec.\ref{sec:gws}  and conclude with a brief summary of our main results in Sec.~\ref{sec:conclusion}. 

\section{Setup and Equation of motion}\label{sec:setup}

Employing Eq.~(\ref{eq:setup}) in Eq.~(\ref{eq:Horndeski}), we obtain
\begin{eqnarray}\label{eq:NMDCandGinf}
S = \int d^{4}x\sqrt{-g} \left[ \frac{M_{pl}^2}{2} R - \frac{1}{2} \left(g^{\mu\nu}-\frac{\alpha}{M^3}\xi(\phi)g^{\mu\nu}\partial_\rho\partial^\rho\phi +\frac{\beta}{M^2}G^{\mu\nu} \right)\partial_\mu\phi\partial_\nu \phi - V(\phi)\right]+S_{m, r}\,,
\end{eqnarray}
where $S_{m, r}$ denotes the standard matter and radiation components.  
The presence of $\xi(\phi)$ and $\beta$ ($\neq0$) makes our model different from the previous study in Ref.~\cite{Brahma:2019kch}.\footnote{When $\xi =-1$ and $\beta=0$, Eq.~(\ref{eq:NMDCandGinf}) is the same as the action in Ref.~\cite{Brahma:2019kch}.} 
By varying Eq.~(\ref{eq:NMDCandGinf})  with respect to spacetime metric $g_{\mu\nu}$, we obtain the Einstein equation 
\begin{eqnarray}\label{eq:Einsteineq}
&&M_{pl}^2 G_{\mu\nu}= T^{m ,r}_{\mu\nu}+ T_{\mu\nu}^{\phi}\,,
\end{eqnarray}
where $T^{m ,r}_{\mu\nu}$ is the energy-momentum tensor of ordinary matter and radiation components and $T_{\mu\nu}^\phi$ is given by
\begin{eqnarray}
T_{\mu\nu}^{\phi} &=& \partial_\mu\phi \partial_\nu\phi-\frac{1}{2}g_{\mu\nu}\left(\partial_\alpha\phi\partial^\alpha\phi+2V\right)\,\nonumber\\
&&+\frac{\alpha}{2M^3}\left[( \xi \nabla_\alpha\phi\nabla^\alpha\phi)_{(\mu}\nabla_{\nu)}\phi - \xi\square\phi \nabla_\mu\phi\nabla_\nu\phi-\frac{1}{2}g_{\mu\nu}(\xi\nabla_\alpha\phi\nabla^\alpha\phi)_{\beta}\nabla^\beta\phi\right] \,\nonumber\\
&&+\frac{\beta}{M^2}\left[-\frac{1}{2}\nabla_\mu\phi \nabla_\nu\phi R + 2\nabla_\alpha\phi \nabla_{(\mu}\phi R^{\alpha}\,_{\nu)}+\nabla^\alpha\phi\nabla^\beta\phi R_{\mu\alpha\nu\beta}+\nabla_\mu\nabla^\alpha\phi\nabla_\nu\nabla_\alpha\phi\right.\\
& &\left.-\nabla_\mu\nabla_\nu\phi\square\phi-\frac12 G_{\mu\nu} \nabla_\alpha\phi \nabla^\alpha\phi+g_{\mu\nu}\left(-\frac{1}{2} \nabla^\alpha\nabla^\beta\phi\nabla_\alpha\nabla_\beta\phi+\frac{1}{2}\left(\square\phi\right)^2-\nabla_\alpha\phi\nabla_\beta\phi R^{\alpha\beta}\right)\right]\,.\nonumber
\end{eqnarray}
Consequently, from Eq.~(\ref{eq:Einsteineq}), we obtain the evolution equation for the scalar field by using the Bianchi identity $\nabla^\mu G_{\mu\nu}=0$ and the conservation law $\nabla^\mu T_{\mu\nu}^{m,r}=0$: 
\begin{eqnarray}
\nabla^{\mu}\,T_{\mu\nu}^\phi=0\,.
\end{eqnarray}

In a spatially flat Friedman-Robertson-Walker universe with metric 
\begin{eqnarray}\label{eq:flatMetric}
ds^2=-dt^2+a(t)^2\delta_{ij}dx^i dx^j\,,
\end{eqnarray} 
where $a(t)$ is a scale factor, gravitational field equations are obtained as
\begin{align}
&3M_{pl}^2 H^2 = \rho_m +\rho_r +\rho_\phi   \,,\label{eq:EE00}\\
& M_{pl}^2\left(2\dot{H}+3H^2\right) = - \frac{1}{3}\rho_r -p_\phi\,,\label{eq:EEii}\\
& \ddot{\phi} +3H\dot{\phi} + V_{,\phi} -\frac{\alpha}{2M^3}\dot{\phi}\left[\ddot{\xi}\dot{\phi}+3\dot{\xi}\ddot{\phi} - 6\xi\dot{\phi} \left( \dot{H} + 3H^2 + 2 H \frac{\ddot{\phi}}{\dot{\phi}} \right) \right]-\frac{3\beta}{M^2} H \dot{\phi}\left( 2\dot{H} +3H^2 +H\frac{\ddot{\phi}}{\dot{\phi}}\right)=0\,,\label{eq:fieldEq}
\end{align}
where $V_{,\phi}\equiv dV/d\phi$, $\rho_{m}$ and $\rho_{r}$ are the energy densities of non-relativistic matter with $p_m=0$ and radiation with $p_r=\rho_r/3$, respectively, and  
\begin{eqnarray}\label{eq:rhophi}
\rho_\phi&=&\frac{1}{2} \dot{\phi}^2 +V(\phi) +\frac{3\alpha}{M^3}H\xi\dot{\phi}^3 \left( 1-\frac{\dot{\xi}}{6H\xi}\right) - \frac{9\beta}{2M^2}\dot{\phi}^2 H^2\,,\\
p_\phi &=& \frac{1}{2}\dot{\phi}^2 - V  -\frac{\alpha}{M^3}\xi\dot{\phi}^3\left(\frac{\ddot{\phi}}{\dot{\phi}} + \frac{\dot{\xi}}{2\xi }\right) +\frac{\beta\dot{\phi}^2}{2M^2}\left(2\dot{H} +3H^2 +4H\frac{\ddot{\phi}}{\dot{\phi}} \right)\,.\label{eq:pphi}
\end{eqnarray}
Eq.~(\ref{eq:fieldEq}) can also be rewritten in terms of $\rho_\phi$ and $p_\phi$ as
\begin{equation}
\dot{\rho_\phi}+3H\left(1+\omega_\phi\right) \rho_\phi = 0\,,
\end{equation}
where $\omega_\phi\equiv p_\phi/\rho_\phi$ is the equation-of-state parameter of the scalar field. The effective equation of state of this system $\omega_{eff}$ is defined as
\begin{eqnarray}\label{eq:omegaphi0}
\omega_{eff} \equiv -1 -\frac{2\dot{H}}{3H^2}\,.
\end{eqnarray}
In the next section, by introducing the dimensionless variables, we rewrite the above equations in the autonomous form, which in turn is useful to analyze the dynamical behavior of the system.

\section{Analyses of the dynamical system}\label{sec:analysis}

In this section, we examine the background evolution of our model by using the so-called dynamical system analysis method. This technique gives a robust description of the cosmic history based on the existence of critical points and their stability, where each of these points corresponds to a different cosmological phase. We introduce the following dimensionless variables to rewrite Eqs.~(\ref{eq:EE00})--(\ref{eq:fieldEq}) into the autonomous form
\begin{eqnarray}
x_1 &=& \frac{\dot{\phi}}{\sqrt{6} M_{pl} H}\,,\label{eq:x1}\\
x_2 &=& \frac{\sqrt{V}}{\sqrt{3}M_{pl} H}\,,\label{eq:x2}\\
x_3 &=& -\frac{6\alpha}{M^3}\xi(\phi)\dot{\phi} H\,,\label{eq:x3}\\
x_4 &=& \frac{9\beta}{M^2}H^2\,,\label{eq:x4}\\
\lambda &=& -M_{pl} \frac{V_{,\phi}}{V}\,,\label{eq:lambd}\\
\sigma &=& -\frac{M_{pl}}{\sqrt{6}} \frac{\xi_{,\phi}}{\xi}\,.\label{eq:sigm}
\end{eqnarray}   
Here, one can see that the presence of both the derivative self-interaction of scalar field and its derivative coupling to gravity is encoded in $\{x_3, \sigma\}$ and $x_4$, respectively. If $x_4=0$ but $\{x_3, \sigma\}\neq0$, the resulting setup describes Einstein gravity only with the derivative self-interaction of the scalar field. On the contrary, a setup with $x_4\neq0$, but $\{x_3,\sigma\}=0$, describes a system wherein gravity is coupled to the scalar field. To take both terms into account, we consider the case where $\{x_3, \sigma$, $x_4\}\neq0$. 
We rewrite Eq.~(\ref{eq:EE00}) in terms of the aforementioned dimensionless variables
\begin{eqnarray}
1=\Omega_m+\Omega_r+\Omega_\phi\,,
\end{eqnarray}
where $\Omega_{m, r} = \rho_{m, r}/(3M_{pl}^2 H^2)$ and $\Omega_\phi = x_1^2\left[1-x_3(1+x_1 \sigma )-x_4\right] +x_2^2$.

By employing a new time variable $N=\ln a$, we  obtain the dynamical equations as follows
\begin{align}
\frac{d \ln x_1}{dN} &= \frac{\ddot{\phi}}{H\dot{\phi}} -\frac{\dot{H}}{H^2}\,,\label{eq:eqforx1} \\
\frac{d \ln x_2}{dN} &=-\left(\sqrt{\frac{3}{2}} x_1 \lambda+\frac{\dot{H}}{H^2} \right)\,, \label{eq:eqforx2}\\
\frac{d \ln x_3}{dN} &= \frac{\dot{\xi}}{H\xi}+\frac{\ddot{\phi}}{H\dot{\phi}}+\frac{\dot{H}}{H^2}\,,\label{eq:eqforx3} \\
\frac{d \ln x_4}{dN} &= 2 \frac{\dot{H}}{H^2}\,,\label{eq:eqforx4}\\
\frac{d \ln \Omega_r}{dN} &= -\left(4+2\frac{\dot{H}}{H^2}\right)\,,\label{eq:eqforOr}\\
\frac{d\lambda}{dN} &= \sqrt{6} x_1 \lambda^2 (1-\Gamma)\,,\label{eq:eqforlambda}\\
\frac{d\sigma}{dN} &= 6 x_1 \sigma^2 (1-\Delta)\,,\label{eq:eqforsigma}
\end{align}
where $\Gamma\equiv V_{,\phi\phi} V/V_{,\phi}^2$, $\Delta\equiv\xi_{,\phi\phi} \xi/\xi_{,\phi}^2$, and
\begin{align}\label{eq:HddH2}
\frac{\dot{H}}{H^2} &= \frac{\sum_{i=1}^{5}N_i(x_1, x_2, x_3, x_4, \Omega_r)}{\sum_{j=1}^{3} D_j(x_1, x_2, x_3, x_4, \Omega_r)} \,,\\
\frac{\ddot{\phi}}{H\dot{\phi}} &= -\frac{\sum_{i=1}^6 K_i(x_1, x_2, x_3, x_4, \Omega_r)}{x_1 \sum_{j=1}^{3} D_j(x_1, x_2, x_3, x_4, \Omega_r) } \,,\label{eq:phiddphiH}\\
\frac{\dot{\xi}}{H\xi}          &=  -6x_1\sigma\,.
\end{align}
The functions $N(x_1, x_2, x_3, x_4, \Omega_r)$, $D(x_1, x_2, x_3, x_4, \Omega_r)$, and $K(x_1, x_2, x_3, x_4, \Omega_r)$ are given by
\begin{align}
N_1 &= -3 x_1^2 \left[2 x_4^2 - 4 x_4 (2 - x_3) + 3 (2 - 4 x_3 + x_3^2)\right]\,,\nonumber\\
N_2 &= 6 x_1^3 x_3 \sigma (9 + x_4 - 3 x_3)\,,\nonumber\\
N_3 &= 6 x_1^4 x_3 \sigma^2 \left[(4 x_4 +3x_3 )\Delta - 6x_3 \right]\,,\\
N_4 &= - 2 (3 - x_4 - 3 x_3) (3 - 3 x_2^2 + \Omega_r)\,,\nonumber\\
N_5 &= x_1 \left[12(3 + \Omega_r - 3x_2^2)x_3 \sigma -  \sqrt{6}x_2^2 \lambda (3x_3+4 x_4) \right]\,,\nonumber
\end{align}
\begin{align}
D_1 &= 4x_1^2 x_4^2\,,\nonumber\\
D_2 &= -4x_4 \left[1 - x_1^2 (1 + x_3) + 2 x_1^3 x_3 \sigma \right]\,,\\
D_3 &= 3 \left[ 4 + x_1^2 x_3^2 - 4 x_3 (1 + 2 x_1 \sigma)\right]\,,\nonumber
\end{align}
\begin{align}
K_1 &=  -6 \sqrt{6} x_2^2 \lambda \,,\nonumber\\
K_2 &= -2 \sqrt{6} x_1^2 x_2^2 x_4 \lambda \,,\nonumber\\
K_3 &= -3 x_1^4 x_3 \sigma (4 x_4 + 3 x_3)\,,\\
K_4 &= 12 x_1^5 x_3 x_4 \sigma^2 \Delta\,,\nonumber\\
K_5 &= 3 x_1^3 \left[ x_4 (8 - x_3) + 3 (1 + 4 \Delta \sigma^2) x_3\right]\,,\nonumber\\
K_6 &= 3 x_1 \left[ 12 - 3 x_3 +\frac{1}{3}(4 x_4 + 3 x_3)\left(\Omega_r- 3x_2^2 \right)\right]\,,\nonumber
\end{align} 
From Eqs.~(\ref{eq:rhophi})--(\ref{eq:pphi}) and (\ref{eq:omegaphi0}), we obtain the equation-of-state parameters as
\begin{align}\label{eq:omegaphi}
\omega_\phi  &= 1 +\frac{1}{3\left[x_2^2+x_1^2\left(1-x_3-x_1x_3\sigma -x_4\right)\right]} \left[\frac{2\dot{H}}{3H^2}x_1^2 x_4+x_1^2\left(4x_4+3x_3\right)\left( 1 +\frac{\ddot{\phi}}{3 H \dot{\phi}}\right) - 6 x_2^2\right]
\,,\\
\omega_{eff} &= -1 - \frac{2\dot{H}}{3H^2} \,.\label{eq:omegaeff}
\end{align}
If $\xi=\text{const.}$, the dynamical evolution of the system is governed by Eqs.~(\ref{eq:eqforx1})--(\ref{eq:eqforlambda}), since of both $\sigma$ and $\Delta$ become zero by definition. Furthermore, if both $V(\phi)$ and $\xi(\phi)$ are exponential functions of $\phi$, the system can be described only by Eqs.~(\ref{eq:eqforx1})--(\ref{eq:eqforOr}). This is because both $\Gamma$ and $\Delta$ are equal to 1 in that case. In the following section, we investigate the latter case for simplicity. 

\section{Explicit models and Numerical results}\label{sec:model}
In this section, we consider
\begin{eqnarray}\label{eq:potandxi}
V(\phi)=V_0 e^{-\lambda \phi/M_{pl}} \qquad \text{and} \qquad \xi(\phi)=\xi_0 e^{-\sqrt{6}\sigma \phi/M_{pl}}\,,
\end{eqnarray} 
where $\lambda$ and $\sigma$ are constants. By taking the second swampland conjecture that demands $M_{pl} |V_{,\phi}|/V\geq c \sim \mathcal{O}(1)$ into account, we consider $\lambda =c \sim \mathcal{O}(1)$ to be positive for our study, but the sign of $\sigma$ can be either positive or negative. As we mentioned above, both $\Gamma$ and $\Delta$ are equal to 1 for our choice of Eq.~(\ref{eq:potandxi}); hence Eqs.~(\ref{eq:eqforx1})--(\ref{eq:eqforOr}) govern the dynamics of the autonomous system. 

In what follows, we analyze the behavior of the dynamical system described by Eqs.~(\ref{eq:eqforx1})--(\ref{eq:eqforOr}). Before that, let us briefly summarize the essentials of the theory of a dynamical system for a one-dimensional system. The central part of analyzing such a theory is to identify all its critical (or fixed) points. The autonomous equation $\dot{x}=f(x)$ is said to have a critical point at $x=x_0$ if and only if $f(x_0)=0$. A critical point $x_0$ is stable (unstable) if all solutions of $x(t)$ are attracted to (repelled by) the critical point. The stability/instability of the fixed point may also be characterized by means of linearization. In the linear stability theory, given a dynamical system $\dot{x}= f(x)$ with a critical point at $x=x_0$, the system is linearized about its critical point $x_0$ by $\mathcal{M}=Df(x_0) = \left( \partial f_i/\partial x_j\right)_{x=x_0}$, where the matrix $\mathcal{M}$ is called the Jacobi matrix. The eigenvalues of $\mathcal{M}$ linearized about the $x_0$ reveal whether the point is stable or unstable. If all the eigenvalues of $\mathcal{M}$ have negative (positive) real parts, trajectories passing nearby $x_0$ are attracted to (repelled by) the critical point, which is then be called \emph{stable} (\emph{unstable}). If all the eigenvalues have non-zero real parts with both positive and negative signs, then the fixed point is called a \emph{saddle} point. Although our system is not one-dimensional, we employ the linear stability theory to discuss the stability/instability of the dynamical system described by Eqs.~(\ref{eq:eqforx1})--(\ref{eq:eqforOr}).

\begin{figure}[h!] \centering
{\includegraphics[width=0.45\textwidth]{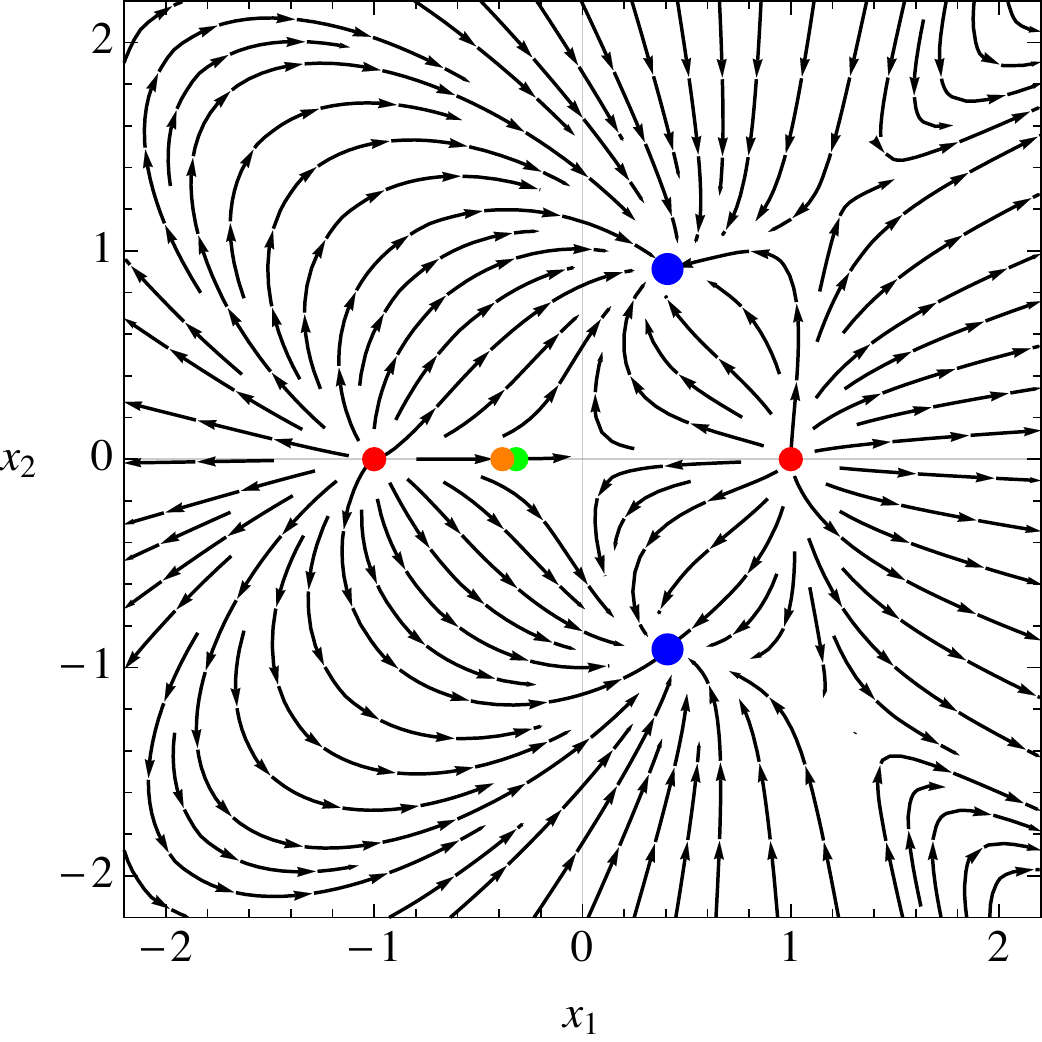}}
{\includegraphics[width=0.45\textwidth]{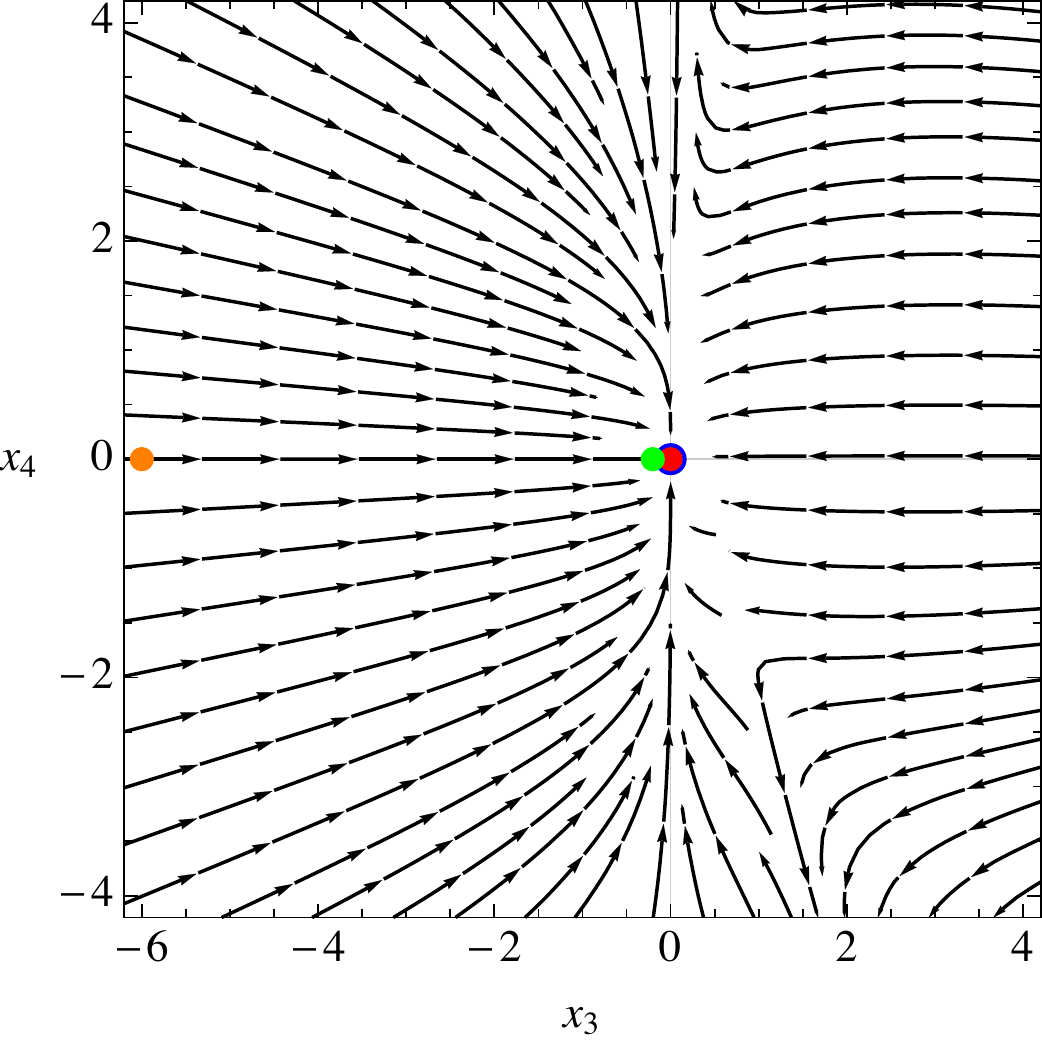}}
\caption{The phase space portraits of the dynamical system Eqs.~(\ref{eq:eqforx1})--(\ref{eq:eqforOr}) with $\lambda=1$ and $\sigma=\sqrt{3}$. The red, blue, green, and orange dots represent the critical $I_{\pm}$, $IV_{\pm}$, $V_\pm$, and $VI$ points in Table~\ref{tab:fxdpnts}, respectively. The critical $II_{\pm}$ and $III_{\pm}$ points are presented due to our choice of $\lambda$ and $\sigma$ values.}\label{fig:stream1}
\end{figure}

Compared to Einstein's gravity, the number of critical points increases due to additional terms considered in Eq.~(\ref{eq:NMDCandGinf}). Depending on the values of $\lambda$ and $\sigma$, we have up to eleven fixed points where $dx_{i}/dN=0$ with $i=(1, 2, 3, 4)$ and $d\Omega_r/dN=0$, which are listed in Table~\ref{tab:fxdpnts}.  Fig.~\ref{fig:stream1} presents the system's dynamical behavior. One can see from Table~\ref{tab:fxdpnts} that the $I_\pm$ points, which denoted by the red dots in Fig.~\ref{fig:stream1}, correspond to solutions where the constraint Eq.~(\ref{eq:EE00}) is dominated by the kinetic energy of the scalar field (\emph{i.e.,} a kinetic regime), and thus the effective equation of state becomes $\omega_\phi=\omega_{eff}=1$. These solutions behave as saddle points, unlike the quintessence case wherein they correspond to unstable nodes and, therefore, are relevant at early times~\cite{Copeland:1997et}.  

The points $II_\pm$ and $III_\pm$ correspond to radiation- and matter-dominated (RD $\&$ MD) phases with $\omega_\phi=\omega_{eff}=1/3$ and $0$, respectively. These points exist only for {\color{black}sufficiently} large values of $\lambda$ (\emph{i.e.,} $\lambda^2>3$ for the RD phase and $\lambda^2>4$ for the MD phase, respectively). Thus, these points are not presented in Fig.~\ref{fig:stream1} because we set $\lambda=1$ to respect the swampland criteria when plotting the figure. 

The blue dots in Fig.~\ref{fig:stream1} denote the fixed $IV_\pm$ points. The system can be stable at these points when certain conditions, as stated in Table~\ref{tab:fxdpnts}, are satisfied for both $\lambda$ and $\sigma$; hence the late-time attractor solutions are possible. These solutions exist for sufficiently flat potential with $\lambda^2<6$. Moreover, Table~\ref{tab:fxdpnts} shows that the equation-of-state parameters for the $IV_{\pm}$ points are proportional to $\lambda^2$ value; therefore, the more we decrease the $\lambda$ value, the more the $\omega_\phi=\omega_{eff}$ value approaches to $-1$. Thus, we find that the late-time acceleration of the universe is possible for these solutions.

The critical $V_\pm$ and $VI$ points, respectively denoted by the green and orange dots in Fig.~\ref{fig:stream1}, correspond to solutions that behave as saddle points. As apparent in Table~\ref{tab:fxdpnts}, the critical $VI$ point, the orange dot in the figure, represents the RD phase with $\omega_\phi=\omega_{eff}=1/3$. However, as Table~\ref{tab:fxdpnts} indicates, these solutions behave as saddle points; hence they cannot give the late-time attractor solutions.

\newpage
\clearpage
\global\pdfpageattr\expandafter{\the\pdfpageattr/Rotate 90}
\begin{turnpage}
\begingroup 
\begin{table}\centering 
\caption{ Fixed points of the autonomous system. \label{tab:fxdpnts}}
\begin{ruledtabular}
\begin{tabular}{ c|c c c c | c c c | c | c | c} 
 Pts. & $x_1$ & $x_2$ & $x_3$ & $x_4$ & $\Omega_r$ & $\Omega_m$ & $\Omega_\phi$ & Existence & Stability/Unstability &  $\omega_{\phi}=\omega_{eff}$ \\ \hline
 $I_{\pm}$ & $\pm 1$ & 0 & 0 & 0 & 0 & 0 & 1 & $\forall \lambda$ & Saddle & 1\\ 
 $II_{\pm}$ & $\frac{2\sqrt{2}}{\sqrt{3}\lambda}$ & $\pm\frac{2}{\sqrt{3}\lambda}$  & 0 & 0 & $1-\frac{4}{\lambda^2}$ & 0 & $\frac{4}{\lambda^2}$ & $\lambda^2>4$ & Saddle & $\frac{1}{3}$\\ 
 \hline
 $III_{\pm}$ & $\frac{\sqrt{3}}{\sqrt{2}\lambda}$ & $\pm\frac{\sqrt{3}}{\sqrt{2}\lambda}$ & 0 & 0 & 0 & $1-\frac{3}{\lambda^2}$ & $\frac{3}{\lambda^2}$ & $\lambda^2>3$ & Saddle for & 0\\ 
 & & &  & & & & & & 
 {\footnotesize $\left(-\sqrt{\frac{24}{7}}\leq \lambda <-\sqrt{3}\land \sigma <-\frac{\lambda }{\sqrt{6}}\right)\lor \left(\sqrt{3}<\lambda \leq \sqrt{\frac{24}{7}}\land \sigma >-\frac{\lambda }{\sqrt{6}}\right)$} \\
  & & &  & & & & & & Stable spiral for \\
  & & &  & & & & & & $\lambda^2>\frac{24}{7}$ and $\forall \sigma$ \\
 \hline
 $IV_{\pm}$ & $\frac{\lambda}{\sqrt{6}}$ & $\pm \sqrt{1-\frac{\lambda^2}{6}}$  & 0 & 0 & 0 & 0 & 1 & $\lambda^2<6$ & Stable for & $-1+\frac{\lambda^2}{3}$ \\ 
 & & &  & & & & & & 
{\footnotesize $ \left(-\sqrt{3}<\lambda <0\land \sigma <-\frac{\lambda }{\sqrt{6}}\right)\lor \left(0<\lambda <\sqrt{3}\land \sigma >-\frac{\lambda }{\sqrt{6}}\right)$} \\ 
 & & &  & & & & & & Saddle for \\
 & & &  & & & & & & $3<\lambda^2<6\land \forall\sigma$  \\
 \hline
 $V_{\pm}$ & $-\sigma \pm \sqrt{\sigma^2-1}$ & 0 & $2-4\sigma \left( \sigma \pm \sqrt{\sigma^2-1}\right)$ & 0 & 0 & 0 & 1 & $\sigma^2>1$ & Saddle & {\footnotesize $-1+2\sigma\left(\sigma\mp\sqrt{\sigma^2-1} \right)$} \\ 
 $VI$ & $-\frac{2}{3\sigma}$ & 0 & $-6$ & 0 & $1-\frac{4}{3\sigma^2}$ & 0 & $\frac{4}{3\sigma^2}$ & $\sigma^2>\frac{4}{3}$ & Saddle & $\frac{1}{3}$\\ 
\end{tabular}
\end{ruledtabular}
\end{table}
\endgroup
\end{turnpage}

\clearpage
\newpage
\global\pdfpageattr\expandafter{\the\pdfpageattr/Rotate 0}

\begin{figure}[h!]\centering
{\includegraphics[width=0.45\textwidth]{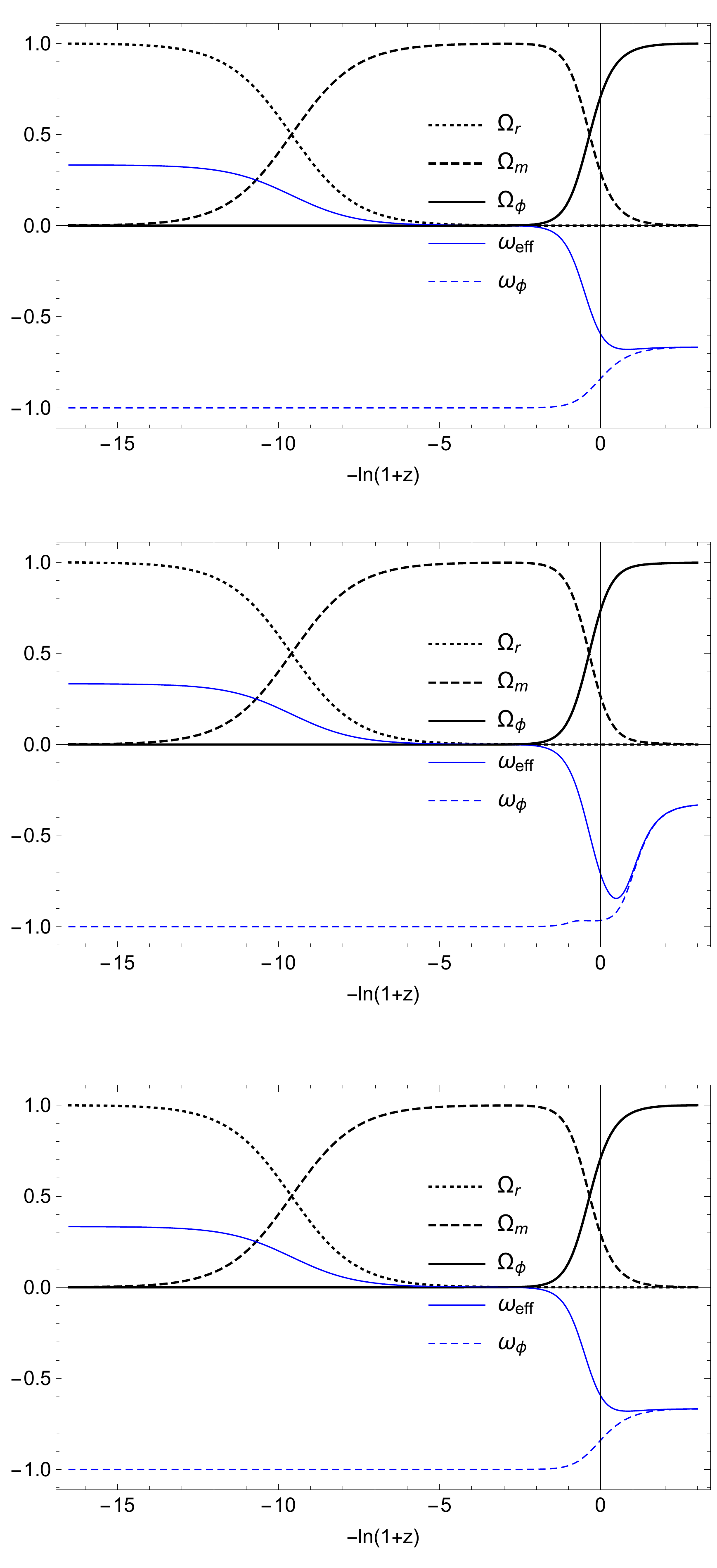}}
{\includegraphics[width=0.45\textwidth]{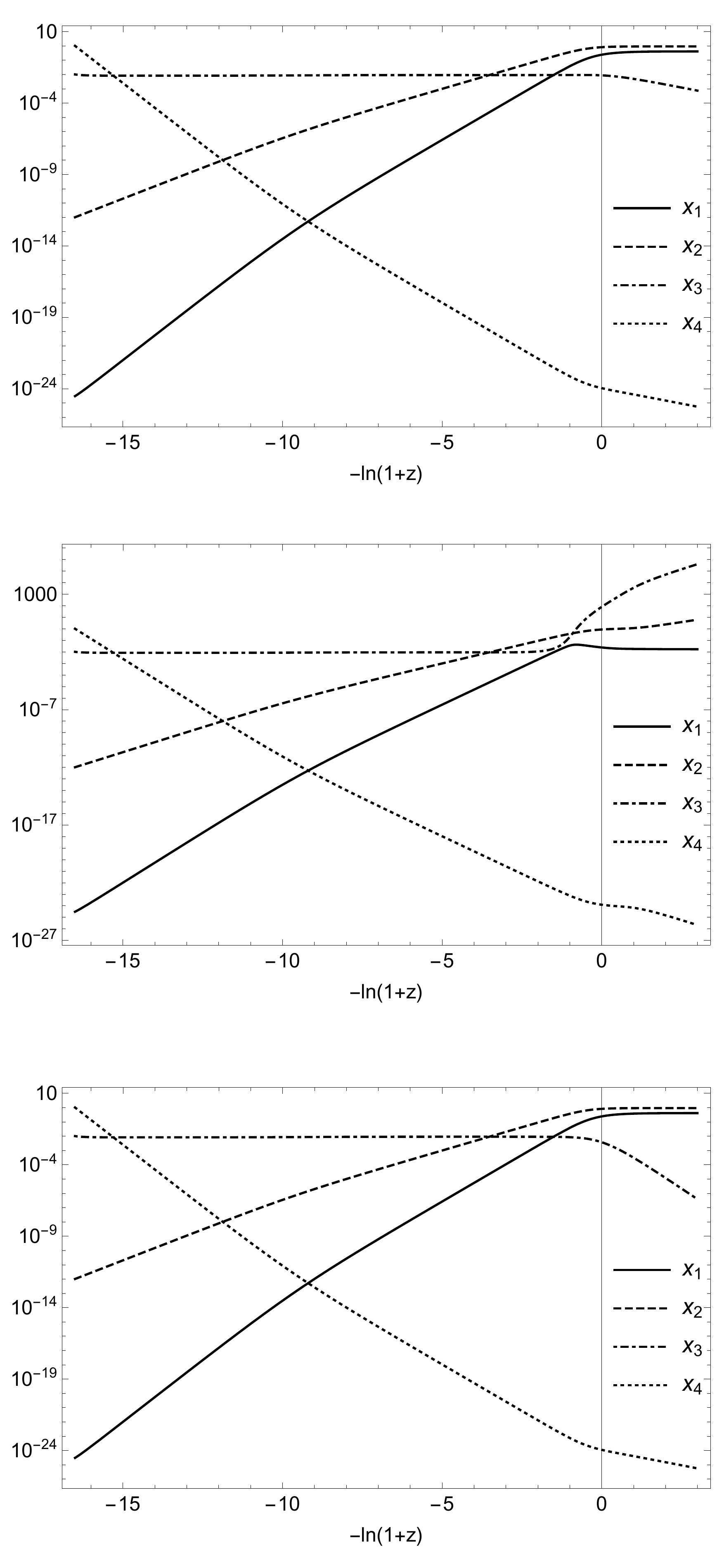}}
\caption{Numerical solutions for $\Omega_i$, $\omega_i$ (left), and $x_i$ (right) with varying $\sigma$, but fixed $\lambda$. Setting $\lambda=1$, we vary $\sigma= \{0,\,-40,\, 1\}$,  from top to bottom panels,  respectively. The initial conditions are given as $x_{1}=3\times10^{-25}$, $x_{2}=10^{-12}$, $x_{3}=10^{-2}$, $x_{4}=1$, $\Omega_{r}=0.999$ at $1+z=1.46\times 10^{7}$.}\label{fig:Omegai}
\end{figure}

Fig.~\ref{fig:Omegai} shows the time evolution of $\Omega_i(z)$ ($i = r, m, \phi$), $x_i(z)$ ($i=1, 2, 3, 4$),  $\omega_{eff}(z)$, and $\omega_\phi(z)$. The initial conditions are given in such a way that the resulting cosmological evolution has the right phase transitions: $\Omega_r$ $\rightarrow$ $\Omega_m$ $\rightarrow$ $\Omega_\phi$. Thus, the effective equation of state starts evolving from $\omega_{eff}\simeq 1/3$ (the RD phase), and after passing through $\omega_{eff} = 0$ (the MD phase), it eventually approaches to $\omega_{eff} \simeq -1$. The universe enters into a phase of the cosmic acceleration when $\omega_{eff} < -1/3$, which occurs around $z\simeq-0.78$ in our case. 

To reproduce a viable cosmic history, the evolution of $x_i$ must depend on fine-tuned initial conditions: $x_4>x_3\gg x_2\gg x_1$ in our case.  Although the initial value of $x_4$ is the largest among them, its time evolution experiences the drastic decrement as the universe expands, see dotted lines in the right column of Fig.~\ref{fig:Omegai}. Such the decreasing behavior can be explained by the time evolution of the Hubble parameter during a phase of accelerated expansion. The estimated value of $x_4\sim \mathcal{O}(10^{-24})$ near $z=0$ leads to a conclusion that the derivative coupling between the scalar field and gravity gets insignificant over time and becomes negligible in the present universe. 

\begin{figure}[h!] \centering
{\includegraphics[width=0.45\textwidth]{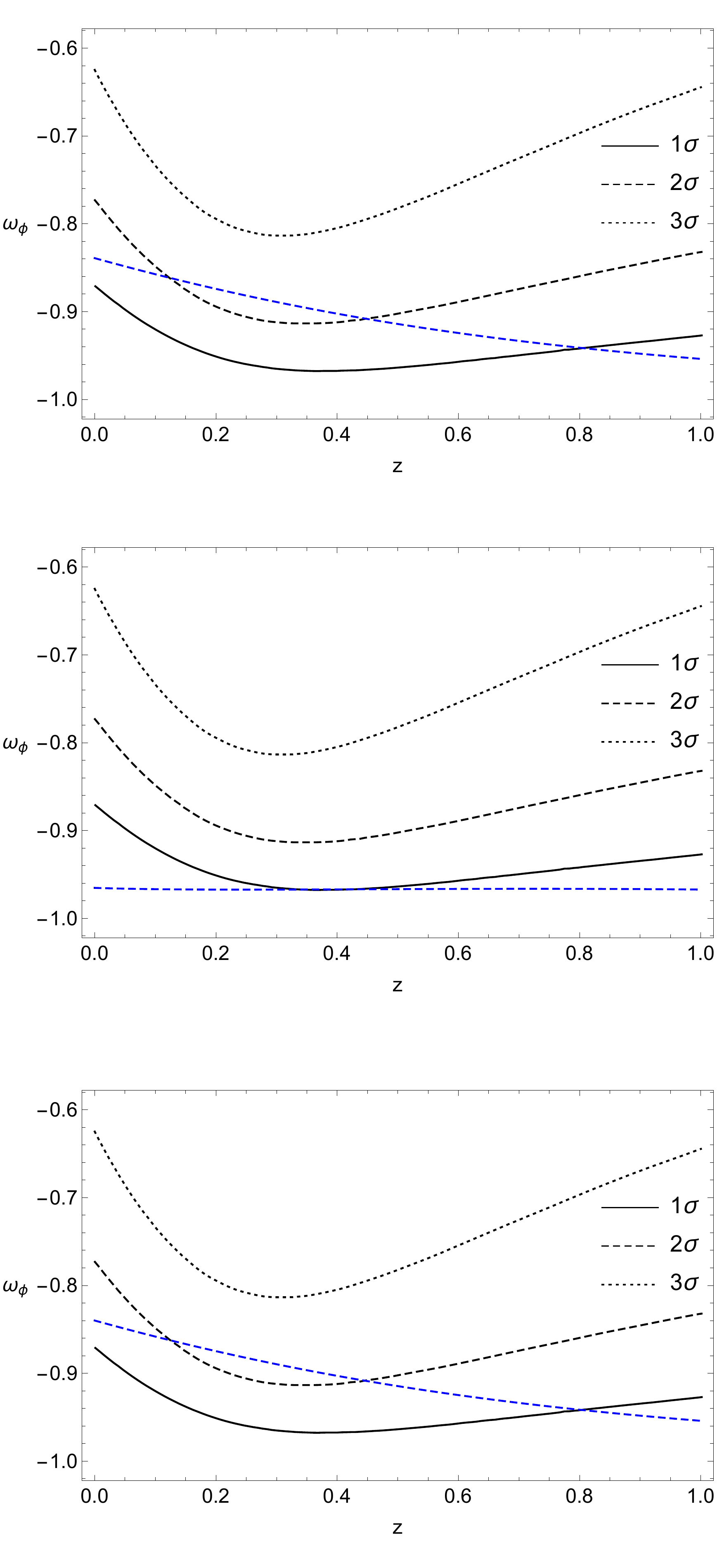}}
{\includegraphics[width=0.45\textwidth]{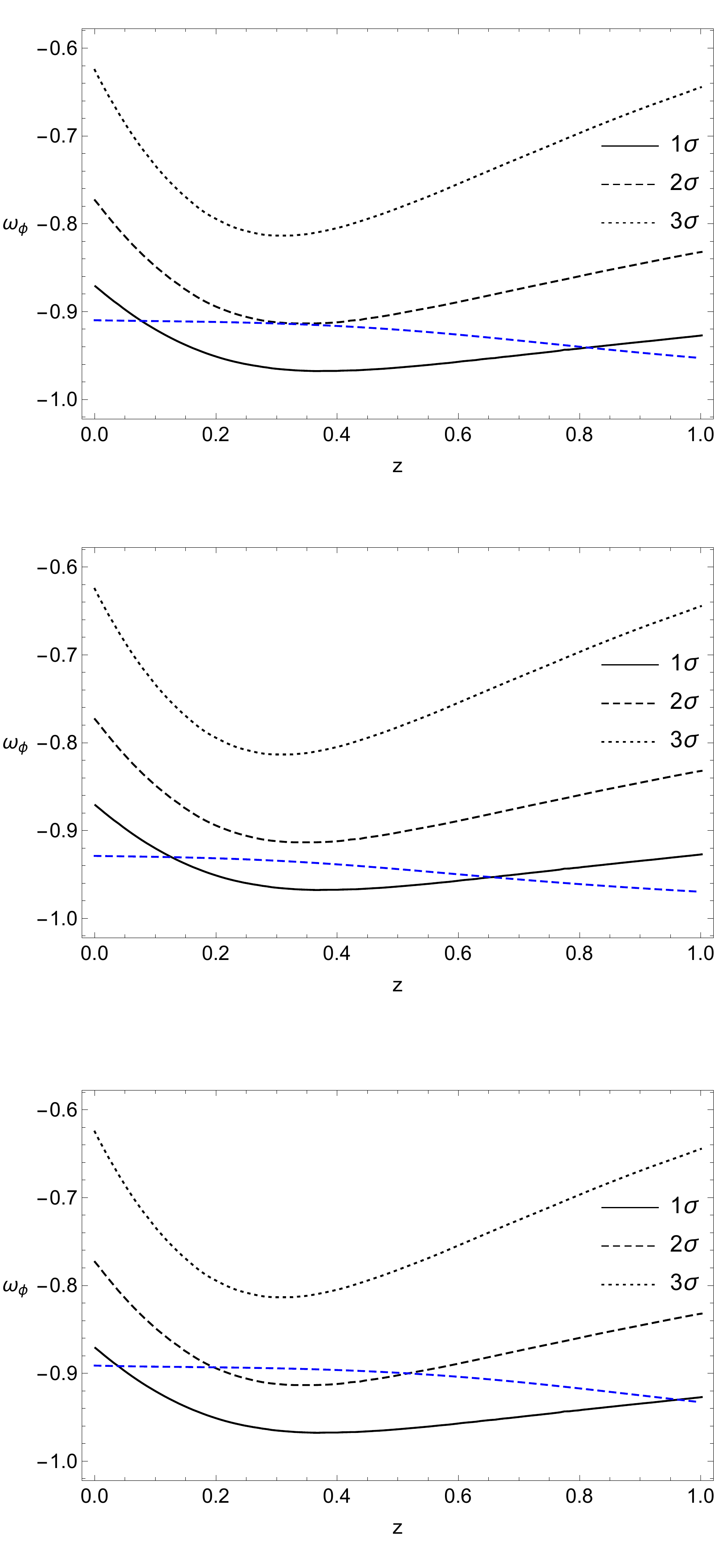}}
\caption{Evolution of $\omega_{\phi}$ with varying $\sigma$ (left) and $\lambda$ (right). The initial conditions are the same as in Fig.~\ref{fig:Omegai}. The blue lines show predictions of our model while the solid-, dashed- and dotted-black ones indicate $1\sigma$, $2\sigma$, and $3\sigma$ contours of the observational upper bounds, respectively. From top to bottom, $\sigma=\{0,\,-40,\, 1\}$ with $\lambda=1$ (left), and $\lambda=\{1,\, 0.8,\, 1.2\}$ with  $\sigma= -14$ (right). }\label{fig:omegai}
\end{figure}

Besides, the time evolution of the derivative self-interaction of the scalar field, \emph{i.e.,} the $x_3$ term, is kept nearly frozen during RD and MD phases, see dot-dashed lines in the right column of Fig.~\ref{fig:Omegai}. However, as the $\Omega_\phi$-dominated era sets in, the $x_3$ eventually increases (decreases) for negative (positive) values of  $\sigma$. In the meantime, for $\sigma\geq0$, $x_1$ and $x_2$ significantly grow and can outpace both $x_4$ and $x_3$ in the vicinity of the MD era. This means the dynamics of our model converge to that of the quintessence in the future. However, for the negative values of $\sigma$ (\emph{i.e.,} $\sigma<0$), the late-time evolution of $x_3$ can grow even faster than $x_1$ and $x_2$, see the dot-dashed line in the middle panel of the right column of Fig.~\ref{fig:Omegai}.~\footnote{For an illustrative purpose, we set $\sigma=-40$ in Figs.~\ref{fig:Omegai} and~\ref{fig:omegai} to emphasize the growth of $x_3$ at late-time for $\sigma\sim\mathcal{O}(1)$ values, which is challenging to notice otherwise.} 
Thus, the dynamical evolution of $\omega_\phi=\omega_{eff}$, approaches to $0$ in the future. This means that in the future, after the scalar-field dominated phase ends, our universe should reenter the MD phase where $\omega_{eff}\simeq0$ once again.

Following Refs.~\cite{Agrawal:2018own, Heisenberg:2018yae, Heisenberg:2019qxz, Brahma:2019kch}, we plot in Fig.~\ref{fig:omegai} the redshift evolution of the equation of state $\omega_{\phi}(z)$ together with observational upper bounds from CMB, BAO, SnIa, and $H_0$ data~\cite{Scolnic:2017caz}.  The theoretical predictions of our model is plotted in blue (dashed) lines with the same initial condition as Fig.~\ref{fig:Omegai}, and we used the Chevallier-Polarski-Linder (CPL) parameterization of the dark-energy equation of state~\cite{Chevallier:2000qy} that reads
\begin{align}
\omega(z) = \omega_0 + \frac{z}{1+z}\omega_a\,.
\end{align} 

The left column of Fig.~\ref{fig:omegai} shows the $\omega_\phi(z)$ with different values of $\sigma$, but fixed $\lambda$, where we set $\lambda=1$ to respect the swampland criteria. We summarize our findings as follows:
\begin{itemize} 
\item[-] When $\sigma=0$, our result in the left-top panel of Fig.~\ref{fig:omegai} reproduces that of the quintessence scenario because $\xi=\text{const.}$ from Eq.~(\ref{eq:potandxi}), and the estimated $x_4$ value in the redshift interval of $0\leq z \leq 1$ is negligible $\lesssim \mathcal{O}(10^{-24})$, see the dotted line in the right-top panel of Fig.~\ref{fig:Omegai}. As a result, the prediction of the model is in conflict with the current observations at the $2\sigma$ level between the redshift interval of $0.1\lesssim z \lesssim 0.45$. 
\item[-] When $\sigma>0$, as apparent in the left-bottom panel of Fig.~\ref{fig:omegai} where $\sigma=1$, the prediction of our model is still in conflict with the current observations at the $2\sigma$ level due to larger values of $\omega_\phi$ within the interval of $0.1\lesssim z \lesssim 0.45$. The redshift interval gets broader as the $\sigma$ value increases. 
\item[-] However, the negative values of $\sigma$ make the theoretical predictions of our model consistent with the current cosmological observations at the $1\sigma$ level, making $\omega_\phi$ approach to $-1$, see the left-middle panel in Fig.~\ref{fig:omegai}. 
\end{itemize}

In the right column of Fig.~\ref{fig:omegai}, we vary the $\lambda$ value by setting $\sigma$ to some negative value (\emph{e.g.,} $\sigma=-14$). Then, respecting the string swampland criteria, we vary the $\lambda$ values between $0.8\leq \lambda\leq1.2$. As apparent in the figure, we obtain a viable result with both the cosmological observations and the swampland criteria. To summarize, Fig.~\ref{fig:omegai} shows that, for $\lambda\sim\mathcal{O}(1)$, the presence of $\xi(\phi)$ with a positive exponent (\emph{or} $\sigma<0$) can make our model more viable and consistent with the current observations and, at the same time, satisfy the  swampland criteria.

\section{Propagation speed of gravitational waves}\label{sec:gws}
\begin{figure}[H] \centering
{\includegraphics[width=0.6\textwidth]{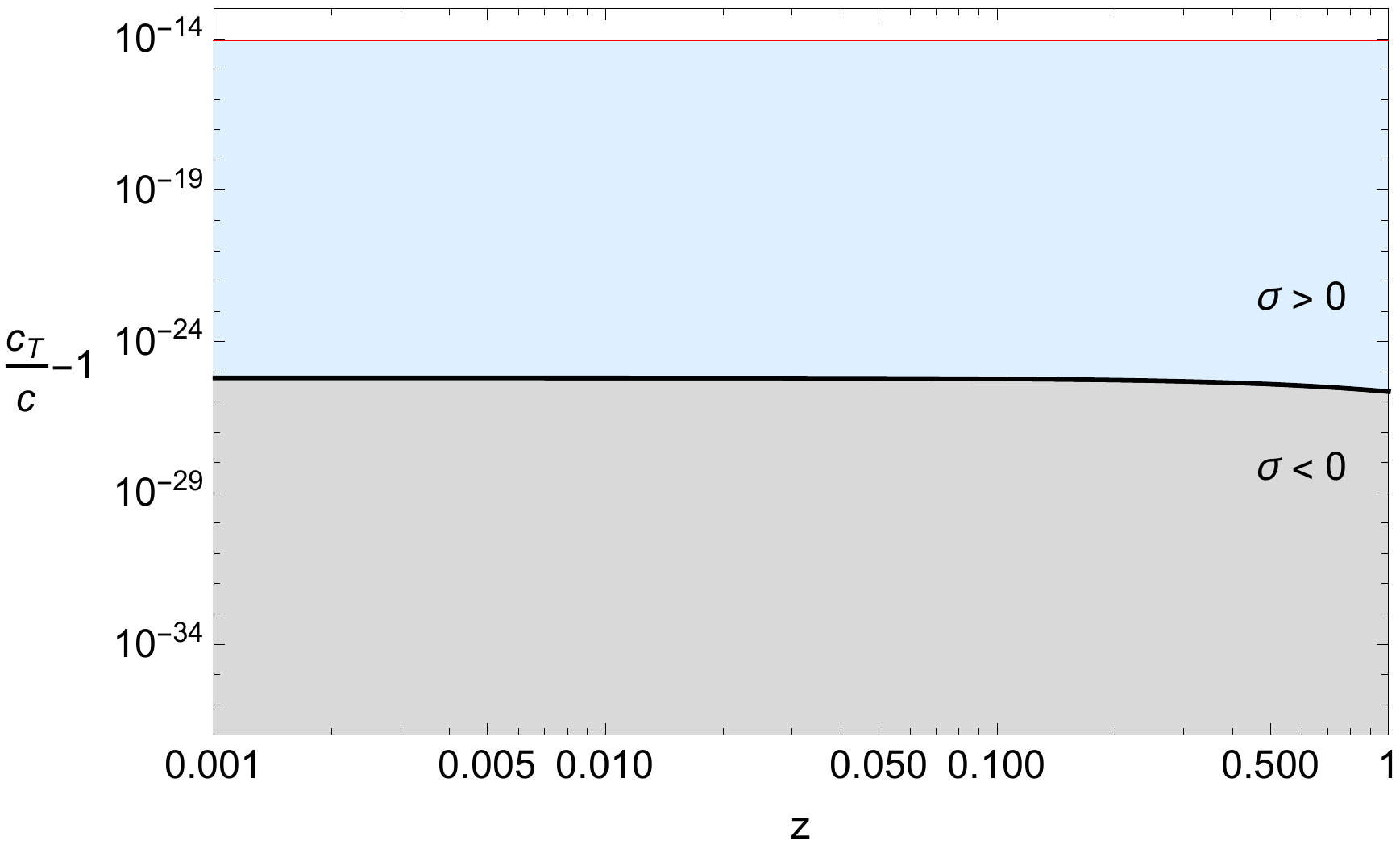}}
\caption{Using the same initial conditions as Fig.~\ref{fig:Omegai}, we plot the $c_{T}/c-1$ where $\lambda=1$. The horizontal red line indicates the observational upper bound from Eq.~(\ref{eq:const-2}). The gray region of $\sigma < 0$ is separated from the cyan one of $\sigma>0$ by the $\sigma=0$ solid line. }\label{fig:ct2}
\end{figure}

In this section, we derive the stringent bounds from the observational constraint on $c_T$.
Substituting our choice of independent function in Eq.~(\ref{eq:setup}) into Eq.~(\ref{eq:ct2}), we rewrite the propagation speed as 
\begin{align}\label{eq:ct2-1}
c_T^2 = \frac{1-\Sigma}{1+\Sigma}\,,
\end{align}
where $\Sigma\equiv \beta \dot{\phi}^2/(2M^2M_{pl}^2) = x_1^2 x_4/3$ in terms of dimensionless variables. Employing the fact that $\Sigma\ll1$ in the late time because $x_4\ll1$ while $x_1^2\sim\mathcal{O}(10)$ from Fig.~\ref{fig:Omegai}, we combine Eq.~(\ref{eq:ct2-1}) with the observational bounds Eq.~(\ref{eq:const}) to obtain
\begin{align}\label{eq:const-1}
-7\times 10^{-16} \lesssim
\Sigma
\lesssim 
3\times 10^{-15}\,,
\end{align}
where $c=1$ assumed. In terms of dimensionless variables, Eq.~(\ref{eq:const-1}) reads
\begin{align}\label{eq:const-2}
-2.1 \times 10^{-15} \lesssim 
x_1^2x_4 
\lesssim 0.9 \times 10^{-14}\,,
\end{align}
which is only valid for our model Eq.~(\ref{eq:NMDCandGinf}). As apparent from Fig.~\ref{fig:ct2}, our model with both positive and negative values of $\sigma$ while $\lambda\sim\mathcal{O}(1)$ is well within the observational bound. The shaded regions (both gray and cyan) in Fig.~\ref{fig:ct2} are favored by the observations.  

\section{Conclusion}\label{sec:conclusion}

In light of the current cosmological observations, we have studied dark energy models from the Horndeski theory of gravity. In particular, we have considered models with the derivative self-interaction of the scalar field and its derivative coupling to gravity. To reproduce the right cosmic evolution ($\Omega_r \rightarrow \Omega_m \rightarrow \Omega_\phi$), we have given initial conditions for $x_i$ as follows: $x_4>x_3\gg x_2\gg x_1$, which indicates that the derivative coupling between the scalar field and gravity must initially be prevailing over terms corresponding to the kinetic and potential energy density, as well as the derivative self-interaction of the scalar field. According to Fig.~\ref{fig:Omegai}, where the time evolution of the coupling between the scalar field and gravity is presented, the effect of such the coupling between the scalar field and gravity gets weaker and weaker over time and eventually becomes negligible in the present universe. 

In Eq.~(\ref{eq:potandxi}), we have chosen the self-interaction term to have an exponential function of $\phi$ with both positive and negative exponents, which correspond to $\sigma<0$ and $\sigma>0$, respectively. For the function that has a positive exponent, our result in Sec.~\ref{sec:model} has shown that the derivative self-interaction term plays an important role in the late-time universe. In other words, we found that while $\lambda\sim\mathcal{O}(1)$, the presence of $ \xi(\phi)\sim e^{-\sqrt{6}\sigma \phi/M_{pl}}$ with a positive exponent (\emph{i.e.,} $\sigma<0$) can make our model more viable and consistent with the current observations and, at the same time, satisfy the swampland criteria. 

We have used the observational bounds on GW speed $c_T^2$ to put constraints on our model in Sec.~\ref{sec:gws}. Fig.~\ref{fig:ct2} has shown that, for a broad range of parameter space, our model satisfies the bounds on the speed of GWs by the GW170817 and GRB170817A measurements. The deviation of the speed of GWs from the speed of light is found to be no more than one part in $\sim 10^{14}$. Our result, at first glance, seems to contradict the conclusion of Ref. \cite{Ezquiaga:2017ekz} in which $G_5 = \text{const.}$ to satisfy the gravitational wave observations. It can be understood as follows: the necessary conditions for the  right cosmic evolution of our model is that the effect of the derivative coupling between the scalar field and gravity term, \emph{i.e.,} $G_5(\phi)$ effect, should be dominant during the radiation-dominated era.  However, $G_5(\phi)$ term decays faster than any other term as time evolves.  As a result, the effect of the $G_5(\phi)$ term at present is negligible.

\begin{acknowledgments}
The authors would like to thank Md. Wali Hossain for his comments, as well as his suggestions, on an earlier version of this paper. BB and ET were supported by Science and Technology Foundation of Mongolia under the contract SHuSS-2019/31. SK was supported by the 2019 scientific promotion program funded by Jeju National University. GT was supported by IBS under the project code IBS-R018-D. SK and GT were supported by the National Research Foundation of Korea (NRF2016R1D1A1B04932574).
\end{acknowledgments}


\end{document}